\documentclass[%
 reprint,
superscriptaddress,
 amsmath,amssymb,
 aps,
]{revtex4-2}

\usepackage{graphicx}
\usepackage{dcolumn}
\usepackage{bm}
\usepackage{hyperref}
\hypersetup{
    colorlinks=true,
    linkcolor=blue,
    filecolor=blue, 
    urlcolor=blue,
    citecolor=blue,
    pdftitle={Overleaf Example},
    pdfpagemode=FullScreen,
    }
\usepackage[mathlines]{lineno}
\usepackage{xcolor}
\usepackage{float}

\begin{document}

\preprint{APS/123-QED}

\title{Polariton-dark exciton interactions in bistable
semiconductor microcavities}

\author{Elena Rozas}
\email{elena.rozas@tu-dortmund.de}
\affiliation{Experimentelle Physik 2, Technische Universit\"{a}t Dortmund, Dortmund D-44221, Germany.}

\author{Evgeny Sedov}
\affiliation{Russian Quantum Center, Skolkovo, Moscow 143025, Russia}
\affiliation{Spin Optics Laboratory, St. Petersburg State University, 198504 St. Petersburg, Russia}
\affiliation{Stoletov Vladimir State University, 600000 Vladimir, Russia}

\author{Yannik Brune}
\affiliation{Experimentelle Physik 2, Technische Universität Dortmund, Dortmund D-44221, Germany.}

\author{Sven Höfling}
\affiliation{Technische Physik, Physikalisches Institut and Würzburg-Dresden Cluster of Excellence ct.qmat, Universität Würzburg, 97074 Würzburg, Germany.}

\author{Alexey Kavokin}
\affiliation{Spin Optics Laboratory, St. Petersburg State University, 198504 St. Petersburg, Russia}
\affiliation{Key Laboratory for Quantum Materials of Zhejiang Province, School of Science, Westlake University, 310024 Hangzhou, China}
\affiliation{Institute of Natural Sciences, Westlake Institute for Advanced Study, 310024 Hangzhou, China}
\affiliation{Moscow Institute of Physics and Technology, Institutskiy per., 9, Dolgoprudnyi, Moscow Region, 141701, Russia}

\author{Marc A\ss{}mann}
\affiliation{Experimentelle Physik 2, Technische Universität Dortmund, Dortmund D-44221, Germany.}

\begin{abstract}

We take advantage of the polariton bistability in semiconductor microcavities to estimate the interaction strength between lower exciton-polariton and dark exciton states. We combine the quasi-resonant excitation of polaritons and the nominally forbidden two-photon excitation (TPE) of dark excitons in a GaAs microcavity. To this end, we use an ultranarrow linewidth cw laser for the TPE process that allows us to determine the energy of dark excitons with high spectral resolution. Our results evidence a sharp drop in the polariton transmission intensity and width of the hysteresis cycle when the TPE process is resonant with the dark exciton energy, highly compromising the bistability of the polariton condensate. This behavior demonstrates the existence of a small symmetry breaking such as that produced by an effective in-plane magnetic field, allowing us to directly excite the dark reservoir. We numerically reproduce the collapse of the hysteresis cycle with the increasing dark exciton population, treating the evolution of a polariton condensate in a one-mode approximation, coupled to the exciton reservoir via polariton-exciton scattering processes.
\end{abstract}

\maketitle

\section{\label{sec:intro}Introduction}

Optically inactive dark excitons have attracted much attention due to their promising prospects for information processing and the creation of long-lasting potentials due to their stable non-radiative nature. For the same reason, applying optical spectroscopy methods for probing them is challenging as one cannot optically excite them and they do not emit light. Thus, dark excitons are not directly accessible by spectroscopy means. To reveal and exploit their properties one has to resort to alternative approaches. For spin-forbidden dark excitons, which we focus on in this manuscript, the use of external in-plane magnetic fields has been examined to induce a mixing of bright and dark states to make spectroscopy experiments possible~\cite{Zhang2017,Lu2019}. This technique enabling redistribution of the weight coefficients between bright and dark states by changing the direction and magnitude of the magnetic field has been studied both experimentally and theoretically in quantum wells (QWs) and quantum dots~\cite{Shamirzaev2017,Shamirzaev2021,Caputo2019a,Sedov2020a,Stevenson2006}. Nevertheless, strong magnetic fields are not routinely available in every laboratory. In addition, strong mixing of dark and bright excitons leads to a significant change in their properties. The presence of a weak symmetry breaking is another option for dark excitons to be accessible via two-photon absorption processes and to allow new transitions. Optical transitions are usually dipole-allowed in one-photon processes and forbidden in two-photon processes, which makes two-photon excitation (TPE) a very selective tool for studying them.

Typically one needs ultrashort pulsed lasers with high peak powers to achieve reasonable TPE efficiencies. This results in a limitation of the achievable spectral resolution. Even worse, at high peak powers second-harmonic generation (SHG) may occur, which may excite bright excitons spectrally close to the dark ones, as in the case of confined systems such as QWs. This is especially common for GaAs QWs, where the energy splitting between dark and bright excitons is very small, of the order of microelectron-volts~\cite{Maialle1993,Vladimirova2010, Trifonov2019}.

In this manuscript, we demonstrate that using an ultra-narrow linewidth high power cw laser together with a polariton system allows us to overcome the indicated issues and yields high-resolution spectroscopic access to dark excitons. For GaAs-based quantum wells embedded inside a high-quality microcavity, polaritons form the bright states and are shifted away from the dark states by the Rabi energy. This increases the bright-dark splitting from a few microelectron-volts for bare quantum wells to about 10 meV for a polariton system, which eliminates the possibility of direct excitation of bright states via SHG. Furthermore, the use of polariton bistability has been already demonstrated as a sensitive probe for tracking the presence and dynamics of dark carriers in semiconductors. A recent experimental study by Schmidt \textit{et al.} found long-lived dark excitons ($>$20 ns) nonresonantly injected into InGaAs quantum wells by monitoring their interaction with lower branch polaritons \cite{Schmidt2019}. Therefore, exciton-polariton systems represent an excellent platform for the observation and study of quantum optical effects. Exciton polaritons are bosonic quasiparticles that stem from the strong coupling of heavy-hole excitons and cavity photons in semiconductor microcavities~\cite{kavokinBook2017}. They inherit unique properties from both their components. The excitonic fraction confers polaritons the ability to interact with each other as well as with excitons and charge carriers \cite{Schmutzler2014b}. The photonic fraction ensures high mobility of polaritons due to the low effective mass and provides direct access to the properties of the system through their photoluminescence. As a result, polaritons have been extensively studied both experimentally and theoretically for their potential applications in quantum optics encouraged by their key role in a wide number of fundamental nonlinear effects including Bose-Einstein condensation \cite{Butov2002,Deng2002Oct,Kasprzak2006,Balili2007, Baumberg2008,Byrnes2014,Sun2017,Su2020,Shan2021,Jiang2022}, polariton lasing \cite{Bajoni2012,Kammann2012,Schneider2013,Kim2016}, superfluidity \cite{Malpuech2007, Utsunomiya2008, Amo2009, Amo2009b,  Amelio2020}, polariton-mediated superconductivity~\cite{Laussy2010,Sedov2020SciRep}, persistent polariton currents~\cite{Sedov2021,Sedov2020} and quantized vortices \cite{Rubo2007,Krizhanovskii2010,Caputo2019,Berger2020}. The origin of these strong nonlinearities lies in the spin-anisotropic polariton–polariton interaction, which is repulsive for like-oriented spins, and manifests itself as a blue shift of the polariton energy~\cite{Whittaker2001}.

In the strong coupling regime, the energy of the bright polariton states is well separated from those of the dark exciton states. Still, despite their optical inactivity, dark excitons may contribute to the wide range of peculiar effects observed in polariton condensates via their interactions with polaritons. In particular, oscillations in the polariton population resembling quantum beats between bright and dark excitons have been observed in the nonlinear regime, which has been explained as a consequence of electron exchange~\cite{Shelykh2005}. 
In order to obtain experimental access to the interactions between dark excitons and polaritons, we take advantage of the bistability exhibited by the polariton photoluminescence when excited in transmission geometry~\cite{Gippius2007, Paraiso2010, Lien2015}. A bistable response can occur when the polariton mode approaches the resonance of the excitation laser, triggering the appearance of a hysteresis cycle between two well-defined states \cite{Baas2004,Schmidt2019, Wouters2013}. The existence of such bistable behavior evidences the presence of an effective internal memory of the polariton system, revealing its prospects for designing memory elements and optical transistors.

In this work, we propose a new approach to quantify the interaction between polaritons and dark excitons based on the analysis of polariton bistability. To do so, we combine the quasi-resonant excitation of lower exciton-polaritons and the TPE of dark excitons. For any system, TPE processes not only require that two photons are absorbed simultaneously, but that those photons arrive at the sample at exactly the same time. To initiate this process, a higher laser power compared to that used for one-photon excitation is required, which is essential for increasing the photon density and, consecutively, the probability of absorption. Sources of ultra-short laser pulses have been widely used for this purpose since they provide a huge number of photons per pulse, i.e., the shorter the pulse, the higher the peak photon flux. However, pulsed lasers necessarily also show a broad energy spectrum which limits the obtainable spectral resolution. To overcome this conundrum, we use the high sensitivity of polariton bistability for detecting weakly populated dark exciton states. This allows us to examine dark excitons using an ultra-narrow linewidth cw laser with high spectral resolution. The possibility of fine-tuning the laser frequency enables us to carefully scan the eigenenergy spectrum of the sample and to identify the dark exciton energy. One should note that the possibility of direct excitation of spin-forbidden carriers even in the absence of an external magnetic field indicates a weak symmetry breaking inherent in our system. 

We observe a significant shift in the hysteresis thresholds introduced by the dark carriers. We reproduce the effect numerically by solving the generalized Gross-Pitaevskii equation, taking into account the condensate energy blueshift introduced by the dark exciton reservoir. The simulations allowed us to estimate the strength of the polariton-exciton interaction responsible for the shift of the thresholds. 

\section{\label{sec:sample}Sample and setup}

The sample used in this work is a planar GaAs $\lambda$ cavity grown by molecular beam epitaxy. Six In$_{0.1}$Ga$_{0.9}$As quantum wells are placed at the antinode positions of the electric field confined by two distributed Bragg reflectors (DBR). The top (bottom) DBR structures consist of 26 (30) pairs of alternating layers of GaAs and AlAs. The microcavity provides a Rabi splitting of polariton eigenmodes of 6~meV. A wedge introduced in the cavity allows modifying the cavity-exciton detuning during the experiments.

\begin{figure}[t]
\includegraphics[width=\columnwidth]{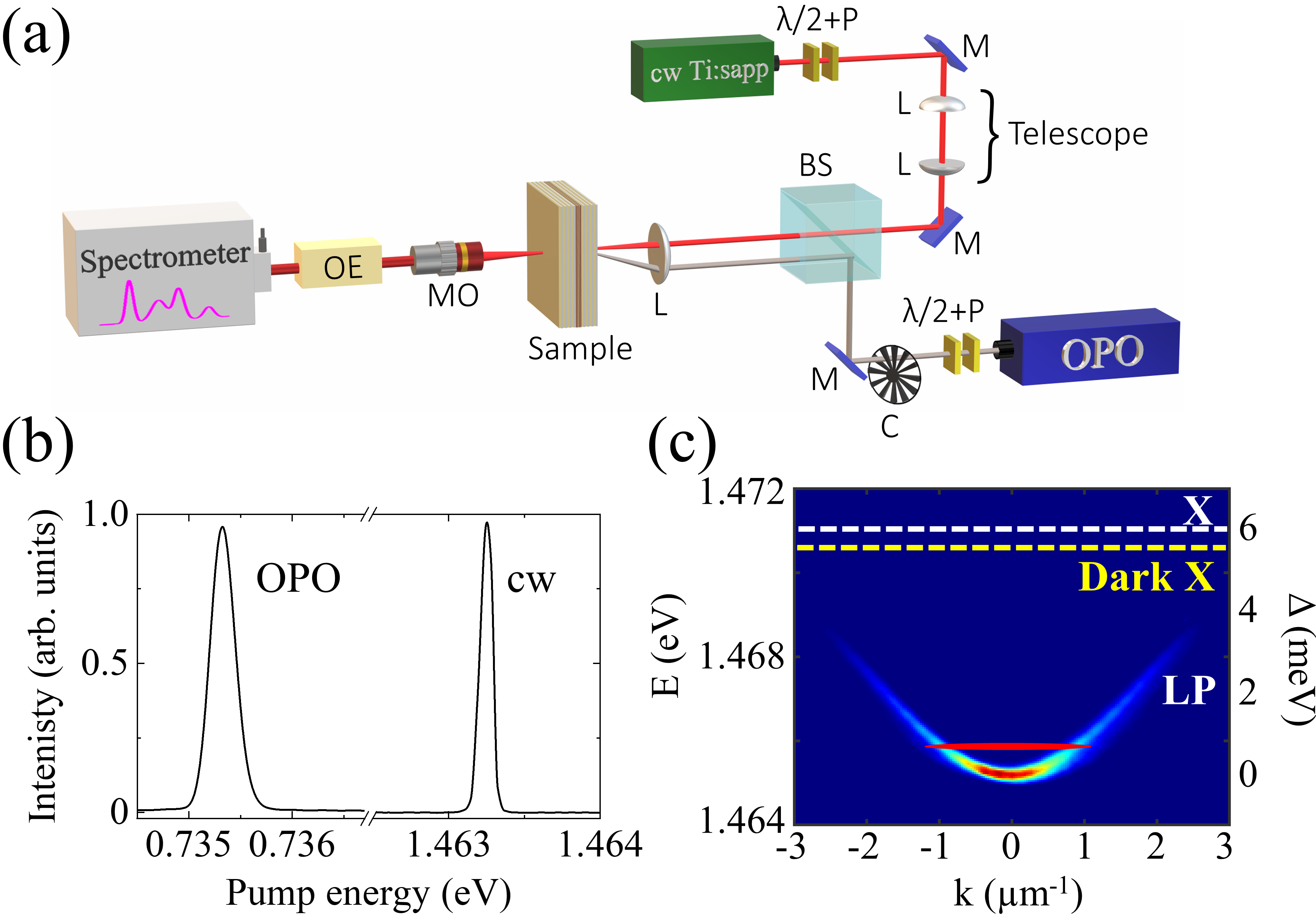}
\caption{\label{Fig1} (a) Sketch of the experimental setup consisting of a continuous wave laser (cw), optical parametric oscillator (OPO), half wave plates ($\lambda$/2), linear polarizers (P), mirrors (M), plano-convex lenses (L), optical chopper system (C), beam splitter (BS), microscope objective (MO) and several optical elements (OE) for the realization of real and Fourier spectroscopy. (b) Normalized laser spectrum of the OPO and cw sources. (c) Angle-resolved photoluminescence of the sample under non-resonant optical pumping. The parabolic dispersion of lower polaritons (LP) is clearly seen. Dashed lines indicate energy levels of bright (X) and dark (Dark X) excitons. The energy of the cw laser is indicated by a red line.}
\end{figure}

A schematic of the experimental setup is displayed in Fig.~\ref{Fig1}(a). The sample was mounted in a helium-flow cryostat at a fixed temperature of 10~K. To create the polariton population, the cavity was excited quasi-resonantly under normal incidence with a cw Ti:Sapphire mode-locked laser detuned by 700~$\mu \text{eV}$ above the LP ground energy state. Additionally, an Aculight Argos MgO:PPLN optical parametric oscillator (OPO) with tunable output wavelength was used for the TPE of dark excitons. In the experiment, the OPO beam was focused onto the sample under an incidence angle of 17$^{\circ}$. Its energy was initially tuned to half the energy of the LP state and gradually shifted towards the exciton level, i.e., $\mathrm{2E_{\text{OPO}}=E_{\text{LP}}+\Delta}$, where $\Delta$ is the energy shift with respect to the initial lower polariton energy. To avoid possible heating effects due to the high excitation power used, the OPO beam was modulated by an optical chopper with a 10$\%$ duty cycle. An example of the energy profiles of both laser beams is shown in Fig.~\ref{Fig1}(b). The measured spectral linewidths are $<$100~$\mu$eV and $<$300~$\mu$eV for the cw and the OPO, respectively. The use of such narrow linewidths is crucial when scanning and determining the energy distribution of dark excitons. Both beams were vertically polarized, parallel to the TM mode of the cavity, and focused onto the sample with a spot size of 28 $\mu$m. An optical telescope was added to the setup to ensure the full overlap between both excitation spots. The light transmitted through the sample was collected at normal incidence using a 10x microscope objective with a numerical aperture of \makebox{NA $=0.26$} and guided to a Si photodiode.

For the correct assessment of the above conditions, the far-field emission from the sample was first collected under non-resonant pumping conditions by tuning the cw source to the first minimum of the DBR stop-band at 1.686~eV. Figure~\ref{Fig1}(c) shows the obtained quasimomentum distribution of lower polaritons with the bottom of the dispersion curve at 1.465~eV. The bare QW exciton level (X) is indicated with a white dashed line at 1.471~eV. The tentative dark exciton level is highlighted with a yellow dashed line at lower energy, which is expected to have a large energy separation from X at $k=0$. The excitation energy of the cw source was later blue-detuned by 700~$\mu \text{eV}$ (red solid line) and the OPO laser was added to the setup. For a better understanding, in the following, we will refer to the OPO energy in terms of the energy shift $\mathrm{\Delta}$ with respect to the unshifted lower polariton energy.

\section{\label{sec:dark} Dark exciton reservoir}

Different mechanisms may contribute to the formation of a dark exciton reservoir, so one should consider which ones are expected to take place in our experiment. First, excitons propagating with large in-plane momenta in the QW plane are optically dark. Since their momenta are beyond the linear photon dispersion (the light cone), they are unable to couple to photons directly. Consequently, the presence of these excitons has been poorly studied because of the inaccessibility of their properties by conventional optical methods. Nonetheless, they have been predicted in low-dimensional semiconductor systems such as QWs and transition metal dichalcogenide monolayers~\cite{Selig2016}. 
The most prominent pathway of their assembly is the spontaneous formation of excitons from free carriers. They are therefore highly relevant for non-resonant excitation of polaritons via excitation above the band gap but are not expected to be relevant for the resonant excitation used in our experiment. Thus, dark excitons within the light cone, where simultaneous conservation of energy and momentum in optical transitions are in principle possible, but orbital or spin quantum number selection rules render one-photon transitions forbidden, are more appropriate candidates in our study. The relevant states that are dark due to orbital quantum number selection rules are given by excited exciton states carrying non-zero orbital angular momentum, such as p--, d--, or f--excitons. In particular, p-exciton states may be excited resonantly via two-photon absorption \cite{Steger2015,Schmutzler2014,Lemenager2014}. However, the 2p-state as the lowest energy p-state exciton is about 10\,meV above the 1s-exciton state \cite{Nithisoontorn1989}, which is above the range of energies we investigate in this study. Therefore, we will also not be able to observe the direct excitation of p-excitons in our experiment.

Another type of dark exciton state is given by excitons with spin-forbidden transitions. Unlike bright excitons, which arise as a result of the interaction of electrons and holes with anti-parallel spins, this type of dark exciton is formed by electrons and holes with parallel spins and acquires the pseudospin $J= 2$ and a spin projection $M_J$. Due to the optical selection rules, they are also optically inactive. As the operators representing electric dipole transitions do not operate on spin space, this is still true in two-photon absorption based on electric dipole transitions. However, two possible excitation pathways remain. Transitions to dark exciton states with $J= 2$ and $M_J= \pm 1$ are allowed due to magnetic dipole two-photon absorption \cite{Michaelis1996}, but very weak. Also, for dark excitons with $J= 2$ and $M_J= \pm 2$, a symmetry breaking may result in mixing between the dark $J= 2$ and the bright $J= 1$ states, enabling the optical excitation of spin-forbidden states and their subsequent interaction with polaritons. This symmetry breaking is commonly introduced by an external magnetic field or by an effective magnetic field, e.g., due to spin-orbit coupling. As we do not apply external magnetic fields, we expect the mixing between dark and bright states to be also weak. In the following, we will show that although these transitions to the spin-forbidden exciton states are only weakly allowed, they have a significant effect on the bistability we observe in the transmission through a polariton microcavity. 

\begin{figure}[t]
\centering
\includegraphics[width=\columnwidth]{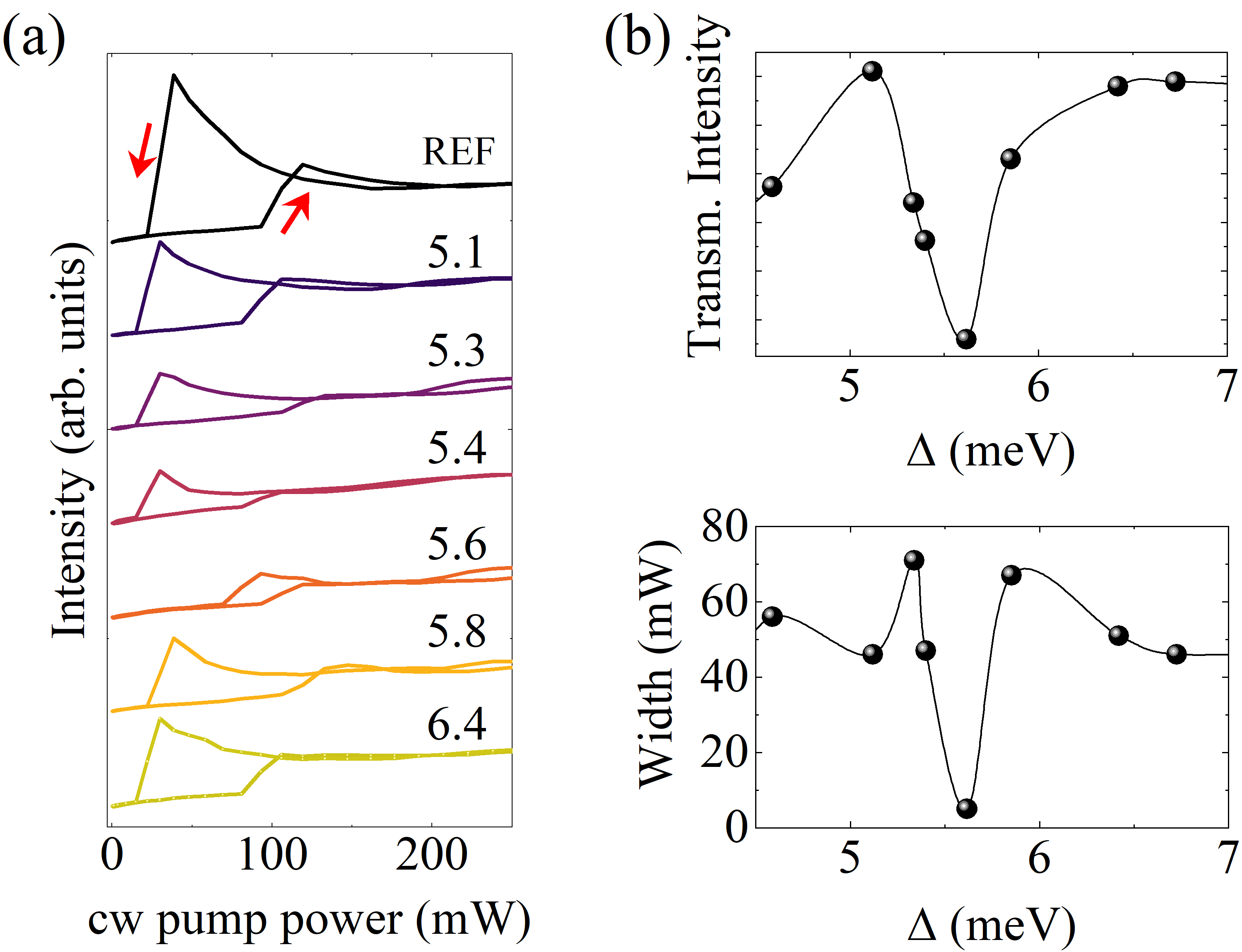}
\caption{\label{Fig2} (a) Transmitted intensity as a function of the cw pump power for different values of $\Delta$ in meV. The hysteresis cycle in the absence of OPO is labeled as REF. For the rest of the profiles, a constant OPO pump power of 1.6 W was used. Red arrows indicate the direction of the tuned power. (b) Transmitted intensity of the ON-state at 70~mW (top) and width of the hysteresis cycle (bottom). Both magnitudes exhibit a dramatic drop at the same value,~$\Delta=$ 5.6 meV.}
\end{figure}

In order to identify the influence of the dark exciton reservoir, in Fig.~\ref{Fig2}(a) we show the transmitted intensity of the polariton photoluminescence. When only the resonant cw pump is applied (REF), near 95~mW the emission indicates a sudden nonlinear increase of the polariton population with increasing pump power. During this transition, the ground mode experiences an energy blueshift as a consequence of the polariton-polariton interaction, shifting the cavity mode closer into resonance with the laser which in turn increases the number of created polaritons and, eventually, the system switches into the ON--state. When the incident pump power is reduced, the cavity--laser resonance remains stable, keeping the system in the ON--state until the power drops below a critical power of 39~mW, which switches the system back to the OFF--state. The region where both states are stable is known as the bistable region. 
It is bounded by two power thresholds that define the switching to the ON-- and OFF--states, which we refer to as the upper and the lower thresholds. Weak changes in the hysteresis loop are observed when the OPO excitation is added to the setup with a pump power of 1.6~W, affecting the transmitted intensity and the thresholds of the bistable region, see Fig.~\ref{Fig2}(a). However, the bistability region considerably reduces when $\Delta$ is tuned close to 5.6~meV. This effect is best illustrated in Fig.~\ref{Fig2}(b), which shows the change of the transmitted intensity in the ON--state and the loop width with ~$\Delta$. The considerable drop in both magnitudes experienced at the same energy indicates the interaction of polaritons with additional carriers emerging in the system. This interaction contributes to the blueshift of the polariton energy and a lower density of polaritons is required to shift the system into the ON--state. One should underline, that the smooth shape of both curves in the vicinity of the minimum can give us an idea about the spectral linewidth of these carriers. 

We attribute this effect to the formation of a spin-forbidden dark exciton reservoir when the OPO is set to $\Delta=5.6$\,meV. Let us stress again that in microcavity structures at $k=0$, the energy splitting between bright and dark exciton states is given by the Rabi splitting and is therefore much larger compared to bare quantum wells, so there are no bright states in the vicinity of the two-photon OPO energy. Hence, we can estimate the energy of dark excitons as $E_{\text{D}} = 1.4708 \pm 0.0003$~eV, which is consistent with the already observed energy splitting between bright and spin-forbidden 1--s exciton dark states in bare QWs \cite{Blackwood1994,Trifonov2019}. The error has been determined by the wavelength resolution of the OPO and the data extracted from the polariton system. Note that heating effects can be safely ruled out as a possible source of the observed effect as heating would be equally noticeable for all OPO energies. Furthermore, we repeated the above procedure for several cavity-exciton detunings (not shown), consistently achieving the same energy for dark excitons in all cases, as one would expect due to the flat dispersion band of excitons.

Although the direct optical excitation of these spin-forbidden 1--s exciton states is at most very weakly allowed in the TPE regime, we also find a small, but characteristic non-linear increase in the total absorption of the OPO light by the sample, when the OPO is tuned exactly to $\Delta=5.6$\,meV (see supplemental material \cite{Suppl}). Accordingly, we have shown that the polariton bistability indeed is a highly sensitive tool for detecting a small population of dark exciton states.

\section{\label{sec:d}Polariton-Dark exciton interaction}

\begin{figure}[t]
\includegraphics[width=0.8\columnwidth]{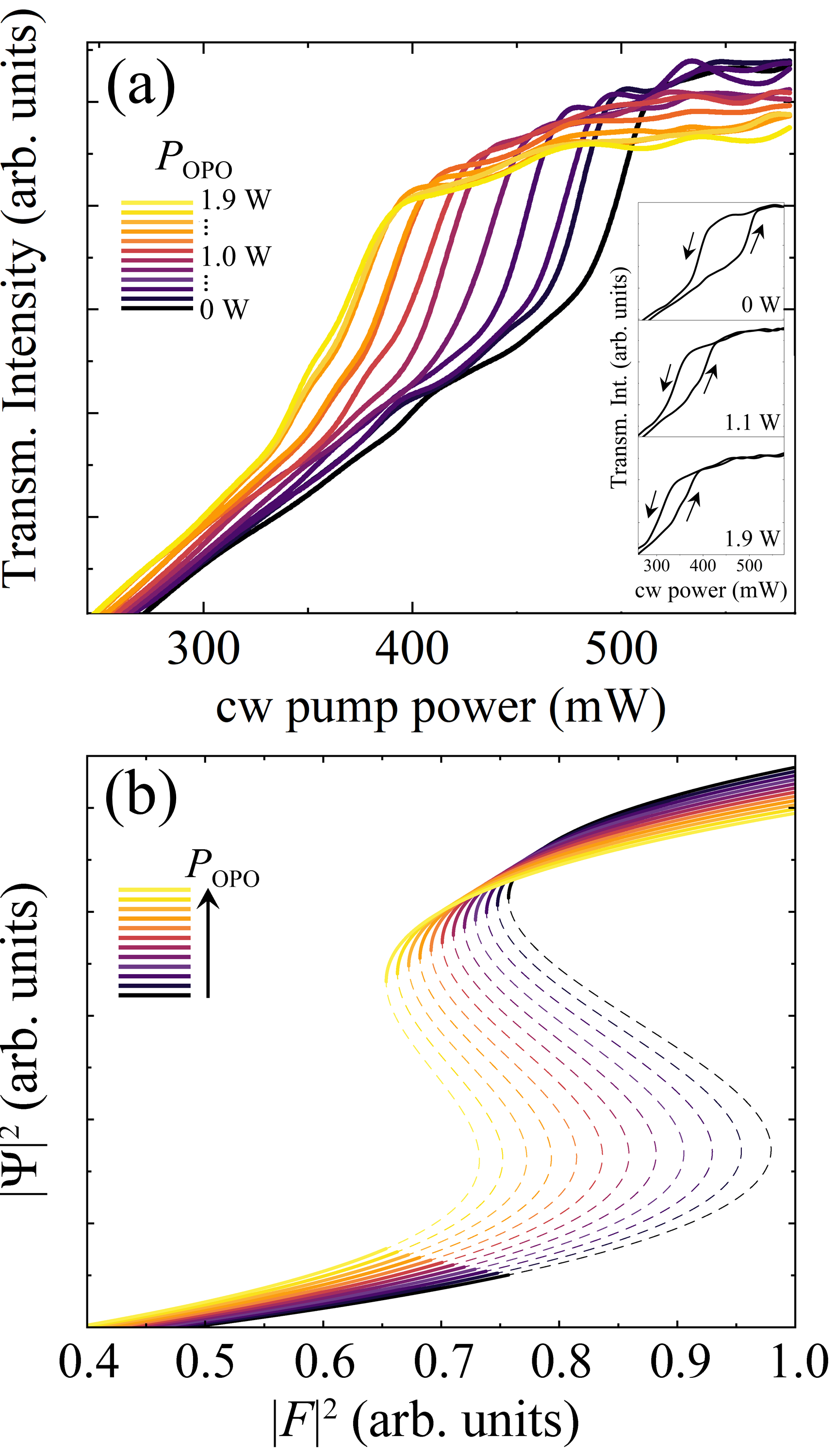}
\caption{\label{Fig3}
Polariton occupation versus increasing cw pump power for different OPO pump powers. (a) Results of experimental observations at $\Delta$ set to 5.6~meV and the OPO pump power ranging between 0 and 1.9~W. The profiles are detected in a region of the sample with $\delta_{\text{C-X}}=0.4$~meV. The inset shows the complete measured hysteresis loops with increasing and decreasing cw pump power for $P_{\text{OPO}}=0,1.1$ and $1.9$ W.
(b) Results of numerical simulations obtained by solving Eq.~\eqref{EqNunbPol}. $P_{\text{OPO}}$ increases for curves from black to yellow. For the simulations, we take $|X|^2 = 0.54$, $\hbar \omega _{\text{p}} = 0.67$~meV and~$\hbar \gamma = 0.276$~meV.}
\end{figure} 

The strength of the interaction between polaritons and dark excitons considerably depends on their population densities, and it can be adjusted by tuning the incident cw and OPO pump power. To evaluate the interaction strength via the formation of the polariton condensate, we investigate the dependence of the transmitted intensity on the cw pump power at different OPO powers, see Fig.~\ref{Fig3}(a). The OPO energy was set to the resonant dark exciton energy level and its power was tuned in the range [0, 1.9]~W. Heating effects arising from the presence of the intense OPO beam were experimentally ruled out by (i) using an optical chopper, (ii) accurately monitoring the lattice temperature, and (iii) by the fact that heating would have affected the experimental data in the same way for all OPO pump energies, which contradicts the results of our observations. It is important to emphasize that chopping the OPO beam is not detrimental to the bistability as the cw laser is still switched on constantly. The interaction with the dark exciton reservoir created by the OPO just shifts the transmission into the ON--state and while the cw power is above the lower threshold, the system stays in the ON--state even when the OPO is not present anymore. The perturbation caused by the presence of the dark reservoir is manifested as a shift of the upper threshold from about 500 to about 350~mW when the OPO power changes from 0 to 1.9~W. More surprisingly, larger dark exciton densities lead to a smoother nonlinear transition between the OFF-- and ON--states, thereby reducing the initially observed slope. Consequently, the bistable region undergoes a significant reduction in its width. As depicted in the inset of Fig.~\ref{Fig3}(a), where the full hysteresis loop is shown for certain $P_{\text{OPO}}$ values, this reduction is also caused by a further displacement of the lower threshold. 

We support the experimental observations with a numerical simulation of the co-evolution of the polariton condensate under the cw pump and the dark reservoir excited by the OPO pump. Following~\cite{Schmidt2019}, we solve a one-mode generalized Gross-Pitaevskii equation for the polariton condensate wave function~$\Psi$, coupled to the rate equation for the occupation number $N_{\text{D}}$ of the dark reservoir:
\begin{subequations}
\label{EqGPEandResEq}
\begin{eqnarray}
&&i d_t \Psi = \left[ -\omega _{\text{p}} + \alpha _{\text{PP}} |\Psi|^2 + \alpha _{\text{PX}} N_{\text{D}} \right] \Psi - i \gamma \Psi + F, \quad \\
&&d_t N_{\text{D}} = W_{\text{OPO}} - \left(\gamma _{\text{D}} + A N_\text{D}{} \right) N_{\text{D}},
\end{eqnarray}
\end{subequations}
where $\omega _{\text{p}}$ is the frequency of the resonant pump with respect to the bottom of the polariton dispersion, and $F$ is the amplitude of the pump.
$\alpha _{\text{XX}}$ is the dark exciton interaction constant which we use as a fitting parameter. In principle, we also need to consider the interaction constants $\alpha _{\text{XX,BD}}$ between bright and dark excitons and $\alpha _{\text{XX,BB}}$ between two bright excitons separately. However, all of them are expected to be at least similar in magnitude \cite{Schindler2008}, so we minimize the number of fitting parameters by defining
$\alpha _{\text{PX}} = |X|^2 \alpha _{\text{XX}} / 2 N_{\text{QW}}$ and 
$\alpha _{\text{PP}} = |X|^2 \alpha _{\text{PX}}$
as the polariton-exciton, and polariton-polariton interaction constants, respectively.
$|X|^2$ determines the exciton fraction in the polariton state and $N_{\text{QW}}$ is the number of QWs in the cavity~\cite{Estrecho2019}.
$\gamma$ and $\gamma _{\text{D}}$ are the polariton and dark reservoir exciton decay rates, respectively. $A$~characterizes the nonlinear losses in the reservoir that may result from, e.g., Auger processes.
$W_{\text{OPO}}$ is the two-photon excitation rate for the reservoir, which is related to the OPO pump power as follows~\cite{Wherrett1984,Pattanaik2016}: $W_{\text{OPO}} = \eta P_{\text{OPO}}^2 / 2 E _{\text{OPO}}$, where $\eta$ is the absorption coefficient and a fitting parameter of the model.

\begin{figure}[t!]
\centering
\includegraphics[width=0.8\columnwidth]{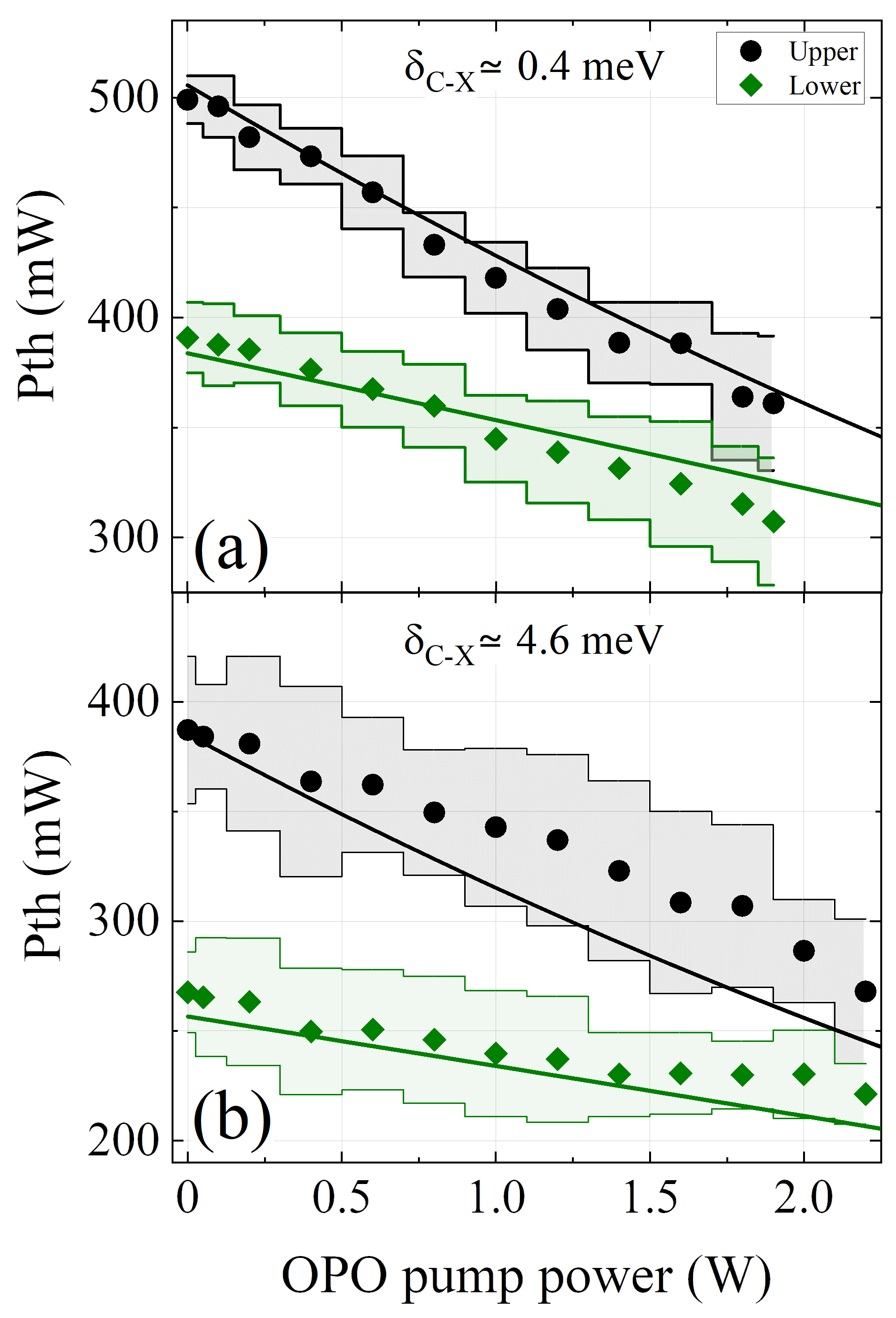}
\caption{\label{Fig4} Upper (black) and lower (green) power thresholds defining the hysteresis cycle as a function of the OPO pump power. (a) and (b) panels correspond to $\delta_{\text{C-X}}$ of 0.4 and 4.6~meV, respectively. Solid lines show the results of the numerical simulations. For the latter, we take (a) $|X|^2 = 0.54$, $\hbar \omega _{\text{p}} = 0.67$~meV, $\hbar \gamma = 0.276$~meV and (b) $|X|^2 = 0.64$, $\hbar \omega _{\text{p}} = 0.7$~meV, $\hbar \gamma = 0.26$~meV.}
\end{figure}

For the driven polariton mode, Eqs.~\eqref{EqGPEandResEq} reduces to:
\begin{equation}
\label{EqNunbPol}
|F|^2 |\Psi| ^{-2} = (\omega _{\text{p}} - \alpha _{\text{PP}} |\Psi|^2 - \alpha _{\text{PX}} N_{\text{D}})^2 + \gamma ^2
\end{equation}
with $N_{\text{D}} = W_{\text{OPO}} / (\gamma _{\text{D}} + A N_{\text{D}})$.
In Fig.~\ref{Fig3}(b) the hysteresis loops on the $(|F|^2, |\Psi|^2)$ plane are shown for different pump powers,~$P_{\text{OPO}}$, obtained by solving Eq.~\eqref{EqNunbPol}. The simulations reproduce both the decrease in threshold pump power and the reduction of the loop width with increasing OPO power.

We now focus on the thresholds of the bistable region. For completeness, we consider two different cavity detunings, 0.4 and 4.6 meV which result in different exciton fractions of the polaritons. The displacement of the upper and lower thresholds with $P_{\textrm{OPO}}$ is shown in Fig.~\ref{Fig4}. The experimental values (black circles and green diamonds) were taken as midpoints of the transitions between the OFF-- and ON--states. Shaded areas indicate the full widths of the transitions. A significant difference of 100~mW in the threshold values for the two detunings can be attributed to the excitonic content of the polaritons. More exciton-like polaritons become heavier (acquire a larger effective mass in the microcavity plane) and less mobile. They are less likely to move away from the injection spot which results in the increase of the density of polaritons trapped by the excitation beam. Thus, the process of polariton condensation due to stimulated scattering proceeds more efficiently.

The simulated dependencies of the thresholds of the hysteresis loop on the OPO pump power are shown in Fig.~\ref{Fig4} with solid lines, supplementing the experimental observation results. The dependencies were calculated by solving the equation $d |F|^2 / d |\Psi|^2 = 0$ for the conditions corresponding to those in the experiments. The dark reservoir decay rate was taken as $\gamma_{\text{D}} = 1/22$~ns$^{-1}$ matching the previous experimental estimations of the parameter in the same sample~\cite{Schmidt2019}. The best fit was achieved for the fitting parameters $A= 4 \times 10^{-6}$~ps$^{-1}$ and $\eta = 10^{-4}$, which gives the estimation of the exciton-exciton interaction constant as $\alpha_{\textrm{XX}} = 61$ $\mu$eV~$\mu \text{m}^2$. Our estimations are further supported by the fact that the polariton-polariton interaction constant then takes the value of $\alpha_{\textrm{PP}} = 1.8$ $\mu$eV~$\mu \text{m}^2$, which is in good agreement with previous estimations in InGaAs QWs~\cite{Rodriguez2016,Munoz-Matutano2019,Delteil2019}, especially when considering that the interaction constant for condensed polaritons is generally lower than for uncondensed ones \cite{Snoke2023}.
Our simulations show that despite the dark nature of the reservoir, its interaction with bright polariton states takes place in the bistable microcavity system with an interaction constant estimated as $\alpha_{\textrm{PX}}\simeq 3$ $\mu$eV~$\mu \text{m}^2$.

\section{\label{sec:con}Conclusions}

In summary, we have developed an approach to accessing properties of spin-forbidden dark excitons in QWs embedded in optical microcavities.
The approach is based on spectroscopic measurements utilizing an ultranarrow linewidth cw laser for the TPE of spin-forbidden dark excitons in the regime of polariton bistability. In the experiment, we studied the peculiar dependence of the photoluminescence of the polariton state on the cw pump power under its interaction with dark excitons. 
The accumulation of long-living optically inactive excitons localized under the pump spot causes a shift of the condensate up to higher energies as well as a reduction of the condensation threshold power. The experimental observations have been reproduced numerically, including the shift of the bistability thresholds and the reduction in width of the bistability loop. Based on a comparison of the results of experimental observations and numerical simulations, we were able to estimate the interaction constant of polaritons and dark excitons with each other and among themselves.
Due to their long lifetimes and robustness, dark excitons are excellent candidates for creating functional potentials in polariton condensates. Knowing the exciton-polariton interaction constant is therefore a decisive factor in the engineering of optical polariton devices and enables their deterministic control. The present work provides a promising step towards polariton-based quantum technologies, such as hybrid qubits with improved performance and longer coherence times compared to traditional qubits based on single-electron states or topological polariton qubits, which are more robust against decoherence and errors due to their non-local, topological nature.

\begin{acknowledgments}
This work was supported by the Deutsche Forschungsgemeinschaft through the International Collaborative Research Centre TRR 160, Grant No.~249492093 (project~B7).
E.S. acknowledges state assignment in the field of scientific activity of the RF Ministry of Science and Higher Education (theme FZUN-2020-0013, state assignment of VlSU). 
Numerical calculations of the hysteresis thresholds were supported by the Russian Science Foundation (Grant No. 19-72-20039).
E.S. and A.K. acknowledge Saint-Petersburg State University (Grant No. 94030557).
A.K. acknowledges the support of Westlake University, Project 041020100118 and Program 2018R01002 funded by the Leading Innovative and Entrepreneur Team Introduction Program of Zhejiang Province of China.
\end{acknowledgments}


\providecommand{\noopsort}[1]{}\providecommand{\singleletter}[1]{#1}%

\end{document}